\documentclass[aps,prl,twocolumn,showpacs,floatfix,superscriptaddress,longbibliography]{revtex4-1}
\usepackage{mathrsfs}
\usepackage{amssymb, amsbsy, amsmath, latexsym, dsfont, array, layout,mathrsfs,color,ulem,bm}
\usepackage[colorlinks,linkcolor=blue,anchorcolor=red,citecolor=blue]{hyperref}
\usepackage{multirow}
\usepackage{float}
\usepackage{threeparttable}
\usepackage{geometry}
\usepackage{ulem}
\newcommand{\ket}[1]{\left|{#1}\right\rangle}
\newcommand{\bra}[1]{\left\langle{#1}\right|}

\usepackage{graphicx}
\usepackage{xspace, stmaryrd, ulem}
\definecolor{delete}{rgb}{1.0, 0.0, 0.0}
\definecolor{edit}{rgb}{0.0, 0.0, 0.9}
\definecolor{comment}{rgb}{0.9, 0.0, 0.0}

\begin{document}

\title{Chiral Entangled-State Generation through Dissipative Quantum Dynamics}

\author{Huixia Gao}
\thanks{These authors contributed equally to this work}
\affiliation{Key Laboratory of Quantum Materials and Devices of Ministry of Education, School of Physics, Southeast University, Nanjing 211189, China}
\author{Konghao Sun}
\thanks{These authors contributed equally to this work}
\affiliation{Beijing National Laboratory for Condensed Matter Physics,
	Institute of Physics, Chinese Academy of Sciences, Beijing 100190, China}
\author{Yiwen Han}
\thanks{These authors contributed equally to this work}
\affiliation{CESQ/ISIS (UMR 7006), CNRS and Universit\'e de Strasbourg, 67000 Strasbourg, France}
\author{Lei Xiao}
\affiliation{Key Laboratory of Quantum Materials and Devices of Ministry of Education, School of Physics, Southeast University, Nanjing 211189, China}
\author{Dengke Qu}
\affiliation{Key Laboratory of Quantum Materials and Devices of Ministry of Education, School of Physics, Southeast University, Nanjing 211189, China}
\author{Kunkun Wang}
\affiliation{School of Physics and Optoelectronic Engineering, Anhui University, Hefei 230601, China}
\author{Xiang Zhan}
\affiliation{Key Laboratory of Quantum Materials and Devices of Ministry of Education, School of Physics, Southeast University, Nanjing 211189, China}
\author{Wei Yi}\email{wyiz@ustc.edu.cn}
\affiliation{Laboratory of Quantum Information, University of Science and Technology of China, Hefei 230026, China}
\affiliation{Anhui Province Key Laboratory of Quantum Network, University of Science and Technology of China, Hefei, 230026, China}
\affiliation{CAS Center For Excellence in Quantum Information and Quantum Physics, Hefei 230026, China}
\affiliation{Hefei National Laboratory, University of Science and Technology of China, Hefei 230088, China}
\author{Peng Xue}\email{gnep.eux@gmail.com}
\affiliation{Key Laboratory of Quantum Materials and Devices of Ministry of Education, School of Physics, Southeast University, Nanjing 211189, China}

\begin{abstract}
Dissipation, though often detrimental to quantum entanglement, can be manipulated for the preparation of entangled states, wherein ingeniously designed quantum jump processes drive the system toward the desired steady state.
Here we venture beyond this paradigm, and demonstrate a new type of entanglement generation in dissipative quantum dynamics. Combining driven-dissipative steady-state engineering and adiabatic passage, we propose a general protocol where the final entangled state depends on the chirality of the evolution path in the parameter space, a scheme that is further extendable to multipartite entanglement.
By simulating the Liouvillian dynamics through the quantum Langevin equation for a pair of photons, we experimentally confirm the noise-resistant chiral preparation of various entangled states with high fidelity and concurrence.
Our work establishes parametric chiral dynamics as a scalable and robust tool for controllable entanglement generation, paving the way for its applications in quantum information.
\end{abstract}

\maketitle

\maketitle

{\it Introduction---}Reservoir engineering for state preparation is an important paradigm in quantum open systems~\cite{PCZ96,VWI09,KLK15,MSO19}. Through the elaborate design of external drive and dissipation, the system, initialized in an arbitrary state, can be driven into a steady state of interest~\cite{CPB93,SM02,RKS12}.
Outstanding examples of reservoir engineering include the preparation of dissipation-induced entanglement~\cite{KBD08,DMK08,KRS11,LGR13,VMC11,LRM14},
topology by dissipation~\cite{DRB11,BBK13,LDW22,WBN20}
as well as driven-dissipative many-body pairing states~\cite{DYD10,YDD12} and scar states~\cite{WYZ24,MGG25}.
For experimental implementations, the realization of the quantum jump processes often relies on stroboscopic simulations~\cite{SMN13,RCB16,SED20}, which are prone to error accumulation, thus limiting the overall evolution time.
Schematically, this can be circumvented by a large enough Liouvillian gap so that the steady state is approached sufficiently fast~\cite{CB13,BCL14,Z15}.
A complementary scenario can be found in state preparation through the conventional adiabatic passage in closed systems, where the time evolution is continuous but needs to be sufficiently long for better adiabaticity and fidelity~\cite{ADL05,VP14,FGG00,YZN24,DXP10}.

In this work, we combine the benefits of both worlds, and experimentally demonstrate the chiral entangled-state generation in the dissipative quantum dynamics of photons.
Our scheme is based on the simultaneous engineering of the Liouvillian and the evolution path in the parameter space, such that through a combination of dissipative steady-state engineering and controlled adiabatic passage, the system is prepared in different entangled states depending on the chirality of the evolution path.
Our experiment is facilitated by the design of an efficient implementation of general non-unitary evolutions for photon pairs, which enables the simulation of Liouvillian dynamics through the quantum Langevin equation.
As we illustrate experimentally, our scheme is scalable to the generation of multipartite entanglement, and robust to dephasing and random perturbations.
Due to the generality and scalability of our scheme,  the results pave the way for practical applications of chiral dynamics for many-body state preparation in quantum open systems.

\begin{figure*}[tbp]
\includegraphics[width=0.75\textwidth]{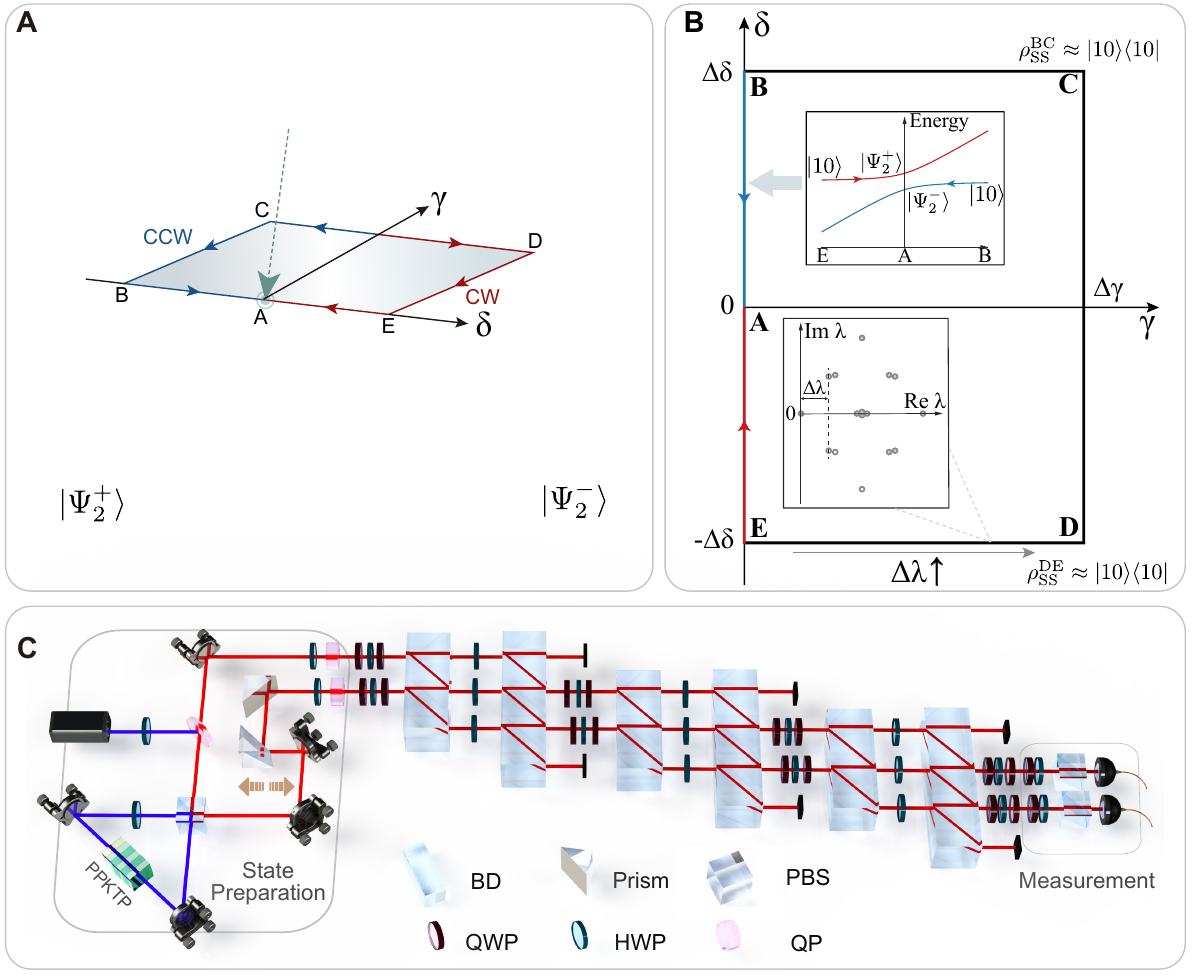}
\caption{Chiral Bell-state generation on a programmable photonic platform.
(a) Conceptual illustration of our protocol. The system is initialized in the maximally mixed state \(\rho(0)=\mathbb{I}/4\). Slow parametric variation along a closed rectangular loop in the \((\gamma,\delta)\) plane prepares different Bell states depending on the chirality of the path: clockwise (CW, red) evolution yields \(|\Psi^{+}_{2}\rangle\), whereas counterclockwise (CCW, blue) evolution yields \(|\Psi^{-}_{2}\rangle\). The tilted rectangle schematically indicates the control loop in the parameter space spanned by \(\gamma\) and \(\delta\).
(b) Encircling path and mechanism. The system evolves along a rectangular path labeled by \(\mathrm{A}\!\mathrm{B}\!\mathrm{C}\!\mathrm{D}\!\mathrm{E}\), with red (blue) arrows indicating the CW (CCW) direction. The upper inset shows the corresponding eigenspectrum of \(H_{0}\) versus \(\delta\).
The lower inset shows a representative Liouvillian spectrum $\lambda$ on the DE segment.
The Liouvillian gap $\Delta \lambda$ increases in the direction of the gray arrow.
(c) Experimental setup. Polarization-entangled photon pairs are generated via type-II spontaneous parametric down-conversion. The desired Liouvillian dynamics are implemented using beam displacers (BDs) and wave plates. At the output, polarization analysis is performed through quantum state tomography.
Detection is carried out with two single-photon avalanche photodiodes, which record two-photon coincidence events.
}
\label{fig1}
\end{figure*}

\begin{figure*}
	\includegraphics[width=0.75\textwidth]{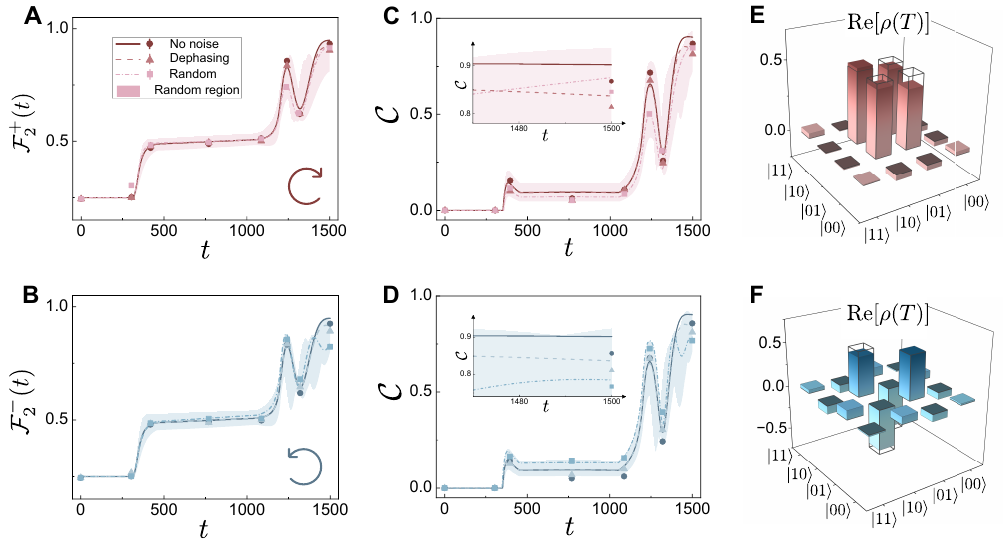}
	\caption{Chiral Bell-state generation.
		(a), (b) Time evolution of the fidelities $\mathcal{F}_{2}^{+}(t)$ and $\mathcal{F}_{2}^{-}(t)$ for the CW and CCW encircling, respectively.
		(c), (d) Time evolution of the concurrence for the CW and CCW encircling, respectively.
		(e), (f) Reconstructed final density matrices obtained from quantum state tomography following the CW and CCW encircling, respectively.
		Parameters: $\xi=1$, $g=0.01$, $\epsilon=1.2$, $T=1500$, $N=200$. Red curves denote the CW evolution, and blue curves denote the CCW evolution. Solid, dashed and dash-dotted curves correspond to results without noise, with dephasing, and with random perturbations, respectively.
		Shaded regions indicate the range of outcomes from $10$ independent realizations of random perturbations.
		Error bars represent statistical deviations from Monte Carlo simulations assuming Poissonian photon-count statistics; some are smaller than the symbol size.}
	\label{fig2}
\end{figure*}

{\it General mechanism and simulation scheme---}We first illustrate our scheme and the underlying mechanism using an interacting two-qubit system, with the Hamiltonian
\begin{align}
	\label{eq.H0}
	H_0=\sum_{j=1,2}\xi_j\sigma_j^{+}\sigma_j^{-}+g\left(\sigma_1^{+}\sigma_2^{-}+\sigma_2^{+}\sigma_1^{-}\right),
\end{align}
where $\sigma^{\pm}_j=(\sigma_x\pm i\sigma_y)/2$ are the raising ($+$) and lowering ($-$) operators of the $j$th qubit, and $\sigma_x$, $\sigma_y$ are the Pauli operators.
The self energies of the two qubits are $\xi_1 = \xi$ and $\xi_2 = \xi + \delta$,
with $\delta$ the detuning in between, and $g$ is their coupling strength.
The qubits are further coupled to the environment, with the dissipative dynamics governed by the
Lindblad master equation~\cite{BP07,HLM17,PKW21}
\begin{align}
	\label{eq.master0}
	\dot{\rho}=\mathcal{L}\rho=-i\left(H_{\text{eff}}\rho-\rho H^{\dagger}_{\text{eff}}\right)+\Gamma\rho\Gamma^{\dagger}.
\end{align}
Here $\rho$ is the density matrix of the two-qubit system, the effective non-Hermitian Hamiltonian
$H_{\text{eff}}=H_0-\frac{i}{2}\Gamma^{\dagger}\Gamma$,
and the quantum jump operator $\Gamma=\sqrt{\gamma}\sigma_1^{+}+\sqrt{\epsilon\gamma}\sigma_2^{-}$,
with the decay rate $\gamma$ and dimensionless ratio $\epsilon$. The Liouvillian gap $\Delta\lambda$, defined as the absolute value
of the real component of the first nonzero eigenvalue of $\mathcal{L}$,
governs the asymptotic relaxation rate toward the steady state. The Liouvillian gap vanishes at $\gamma = 0$, where the dynamics becomes unitary, and continuously increases with larger $\gamma$.

By design, the system has several key properties. First, the product state $|10\rangle$ is a dark state with $\Gamma|10\rangle=0$ for finite $\gamma$, and, in the limit of infinite detuning $|\delta|$, it is also an eigenstate of $H_0$. Hence, $|10\rangle$ is close to the steady state of the dissipative dynamics for sufficiently large $|\delta|$.
Second, in the special case of $\delta=0$, the Bell states $\ket{\Psi^{\pm}_{2}}=(\ket{01}\pm\ket{10})/\sqrt{2}$ are eigenstates of $H_0$.

Figure~\ref{fig1}(a) summarizes the central results of our work. Starting from the maximally mixed state $\rho(0)=\mathbb{I}/4$, a slow closed-loop variation of the system parameters in the $(\gamma,\delta)$ plane prepares different Bell states depending solely on the chirality of the loop: clockwise (CW) encircling drives the system to $|\Psi^{+}_{2}\rangle$, whereas counterclockwise (CCW) encircling drives it to $|\Psi^{-}_{2}\rangle$. While the chiral entanglement generation is apparently similar to chiral state transfer close to exceptional points (EPs), our scheme is based on more general mechanisms that do not necessarily involve EPs.

Take for example the CW encircling path $\mathrm{A}\to\mathrm{B}\to\mathrm{C}\to\mathrm{D}\to\mathrm{E}\to\mathrm{A}$ in Fig.~\ref{fig1}(b), where $2\Delta\delta$ and $\Delta\gamma$ denote the modulation range of $\delta(t)$ and $\gamma(t)$, respectively. Since the Liouvillian gap $\Delta \lambda$ only vanishes along the Hermitian sector $\mathrm{E}\mathrm{A}\mathrm{B}$ (see inset), by the time the system reaches point $\mathrm{D}$, it is already driven to the steady state, provided that the encircling speed is not too fast.
The system would remain in the instantaneous steady state along the path as it approaches point $\mathrm{E}$. For sufficiently large
$\Delta \delta$, the steady state is very close to the product state $|10\rangle$, which is also close to the eigenstate of $H_0$ that would be adiabatically connected to the Bell state $|\Psi^{+}_{2}\rangle$ along $\mathrm{E}\to\mathrm{A}$ (see inset red curve). Similarly, a CCW encircling would drive the system sufficiently close to the eigenstate of $H_0$ at $\mathrm{B}$ that is adiabatically connected to a different Bell state $|\Psi^{-}_{2}\rangle$ (see inset blue curve) on returning to point $\mathrm{A}$. The scheme relies on a careful design of $H_0$, the quantum jump processes, and the encircling path.
Particularly, the Hermitian sector should be designed such that the steady state has a large overlap with the desired eigenstate of $H_0$ at the two ends.

For experimental simulation of the Liouvillian dynamics, we adopt an
equivalent formulation  based on the quantum Langevin equation~\cite{KG92}
\begin{align}
	\label{eq.langvein}
	\dot{\rho}_m(t)=-i\left[\Tilde{H}(t)\rho_m(t)-\rho_m(t)\Tilde{H}^{\dagger}(t)\right].
\end{align}
Here $\Tilde{H}(t)=H_{\text{eff}}+i l(t) \Gamma$, $l(t)$ is the white noise characterized by $\langle l(t)\rangle=0$ and $\quad\left\langle l(t) l^{*}\left(t^{\prime}\right)\right\rangle=\delta\left(t-t^{\prime}\right)$. The symbol  $\langle\cdot\rangle$ represents  the ensemble average.

We discretize the total evolution time $\tau$ into $N$ steps and construct the time-evolution operator $U(\tau)$  in a stroboscopic fashion~\cite{XQW21}, with
\begin{align}
	\label{eq.U}
	U(\tau)=\prod_{k=1}^{N}e^{-i\Tilde{H}(t_k)\delta t},
\end{align}
where $t_k=k\delta t$, and $\delta t=\tau/N$.
Each realization of the noise $l(t)$
generates a distinct quantum trajectory $\rho_m(\tau)=U(\tau)\rho(0)U^{\dagger}(\tau)$, and the overall system density matrix $\rho(\tau)$ is reconstructed by ensemble averaging over all trajectories
\begin{align}
	\label{eq.rho}
	\rho\left(\tau\right)= \frac{1}{n}\sum_{m=1}^n \rho_m\left(\tau\right).
\end{align}

{\it Chiral Bell-state generation---}We first experimentally demonstrate the chiral Bell-state generation.
As illustrated in Fig.~\ref{fig1}(c), our experimental implementation comprises three steps: state preparation, evolution, and quantum state tomography.

For state preparation, the photon pairs are initialized as the maximally mixed state $\mathbb{I}/4$.
The qubits are encoded in the polarizations of the photon pairs, with $\{\ket{00}=\ket{HH}, \ket{01}=\ket{HV}, \ket{10}=\ket{VH}, \ket{11}=\ket{VV}\}$, where $H$ ($V$) denotes horizontal (vertical) polarization.

A key challenge to our experiment is the implementation of the non-unitary evolutions of Eq.~(\ref{eq.U}), especially for the multi-qubit case.
This is overcome by decomposing the evolution operator into tensor products of smaller matrices (see Sec. S2 of the Supplemental Material).
In the current two-qubit case, we have $U(\tau)= Q_{11}\otimes Q_{22}+U_{\text{SWAP}}(Q_{12}\otimes Q_{21})$,
		where $Q_{ij}$ is a $2\times2$ complex matrix, and the index $i$ ($j$) labels the operator acting on the modes of the $i$th ($j$th) photon. The operator $U_{\text{SWAP}}$ is the standard two-qubit SWAP gate.
		The decomposition includes both uncontrolled operations (the first term) and controlled operations (the second term, involving the SWAP gate), allowing for a flexible implementation with our experimental setup~\cite{ZWX20} (see Sec. S2 of the Supplemental Material).

		At the end of the time evolution, we reconstruct the density matrix $\rho_m(t)$ via the two-qubit quantum state tomography. This procedure is repeated for $10$ independently generated noise realizations, each followed by state tomography. The density matrix $\rho(t)$ is then obtained by statistically averaging the reconstructed $\rho_m(t)$ over all realizations, following Eq.~(\ref{eq.rho}).  We then characterize the resulting dynamics by evaluating the fidelities $\mathcal{F}^{\pm}_{2}(t)$ with respect to the Bell states $\ket{\Psi^{\pm}_{2}}$ and the concurrence $\mathcal{C}$~\cite{W98} of the generated entanglement (see Appendix B for details).

		As shown in Figs.~\ref{fig2}(a) and \ref{fig2}(b), CW and CCW encirclings deterministically steer the system toward the two orthogonal Bell states, respectively, revealing the intrinsic chirality of the dissipative dynamics. The concurrence (Figs.~\ref{fig2}(c) and \ref{fig2}(d)) increases from zero at $t=0$ and reaches its maximum at the final time $t=T$, indicating the buildup of strong entanglement during the evolution. The final concurrence reaches $\mathcal{C}=0.8677\pm0.0008$ for CW encircling and $\mathcal{C}=0.8569\pm0.0026$ for CCW encircling, significantly exceeding the maximum concurrence $\mathcal{C}_{\mathrm{max}}^{\mathrm{aut}}\approx0.31$ achievable in a fully autonomous two-qubit system without external drive or control~\cite{P23,BCH22}.
		
		In Figs.~\ref{fig2}(e) and \ref{fig2}(f), we show the final-state density matrix constructed using quantum state tomography.
		Under the CW encircling, the system evolves into the Bell state $\ket{\Psi^{+}_{2}}$ with a fidelity of $\mathcal{F}^+_2(T)=0.9336 \pm 0.0005$, whereas CCW encircling prepares the Bell state $\ket{\Psi^{-}_{2}}$ with a fidelity of $\mathcal{F}^-_2(T)=0.9253 \pm 0.0010$.
		The reconstructed density matrices (colored bars) show excellent agreement with the corresponding ideal Bell states (black open bars).
		These results demonstrate that chiral dynamics provides an efficient route to high-quality entanglement, with the final state uniquely determined by the chirality of encircling.
		
	{\it  Robustness of chiral Bell-state generation---}To test the robustness of the process above, we consider two representative noise sources. We first examine the effect of dephasing~\cite{SY23,GSQ25}. In the presence of dephasing, the system evolution is governed by
		\begin{align}
			\dot{\rho}=\mathcal{L}_{2}^{d}\rho
			=\mathcal{L}\rho
			+\sum_{i=1}^{2}\left(\Gamma_i\rho\Gamma_i^{\dagger}-\frac{1}{2}\{\Gamma_i^{\dagger}\Gamma_i,\rho\}\right),
		\end{align}
		where dephasing is introduced through the jump operators
		$\Gamma_1=0.01\,\sigma_1^z$,
		$\Gamma_2=0.01\,\sigma_2^z$,
		with $\sigma_1^z=\sigma_z\otimes\mathbb{I}$ and $\sigma_2^z=\mathbb{I}\otimes\sigma_z$.
		
		The resulting dynamics are shown as dashed lines in Figs.~\ref{fig2}(a)–\ref{fig2}(d), with all other parameters identical to those in the noiseless case.
		As highlighted by the enlarged views in Figs.~\ref{fig2}(c) and \ref{fig2}(d), the final concurrence is only slightly reduced compared with the noiseless case. Our chiral entanglement generation scheme is thus robust against dephasing.
		
		We then examine the robustness of our mechanism against random perturbations, following the model developed in Ref.~\cite{PLK20}. The ideal Hamiltonian in Eq.~(\ref{eq.H0}) is assumed to be weakly perturbed such that the total Hamiltonian becomes
		$H_0^r = H_0 + H^{r}_{2}$,
		where
		$H^{r}_{2} = 0.01\,\frac{A_{2} + A_{2}^{\dagger}}{2}$.
		Here $A_{2}$ is a $4\times4$ random complex matrix  whose entries are independently sampled as $a+ib$, with $a$ and $b$ uniformly distributed in $[-1,1]$.
		The symmetrization $(A_{2} + A_{2}^{\dagger})/2$ ensures that $H^{r}_{2}$ is Hermitian, while the prefactor $0.01$ sets the perturbation strength.
		
		The dash-dotted curves in Figs.~\ref{fig2}(a)–\ref{fig2}(d) represent the effect of this perturbation.
		The enlarged views in Figs.~\ref{fig2}(c) and \ref{fig2}(d) show only a slight reduction in concurrence relative to the noiseless case.
		The shaded regions indicate the range of outcomes from $10$ independent realizations of random perturbations. In all cases, the final-state fidelity and concurrence remain high.
		
		\begin{figure}
			\includegraphics[width=0.45\textwidth]{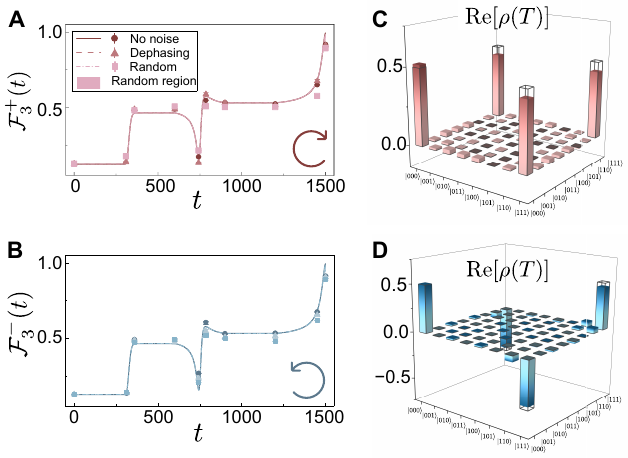}
			\caption{ Chiral preparation of multi-qubit entanglement.
				(a), (b) The fidelities between the evolved states and the target entangled states are shown for CW and CCW encircling, respectively. (c), (d) Reconstructed final density matrices obtained from quantum state tomography following the CW and CCW encircling, respectively.
				The encircling time is $T=1500$, the initial state is $\mathbb{I}/8$, and the parameter $J_x=1$.}
			\label{fig3}
		\end{figure}

{\it  Chiral dynamics in the multi-qubit case---}Our scheme can be generalized to multi-qubit entanglement (see more details in Secs. S1 and S5 of the Supplemental Material).
Here we demonstrate the case of three-qubit entanglement as an example.
For the experimental simulation, we encode the basis states in the hybrid polarization–spatial modes of the photon pairs (see Sec. S2 of the Supplemental Material).
We employ decompositions of the evolution operator similar to the two-qubit case (see Sec. S2 of the Supplemental Material).
To characterize the resulting dynamics, we perform quantum state tomography~\cite{EKB04} and evaluate the fidelities $\mathcal{F}^{\pm}_{3}(t)$ with respect to the target states $\ket{\Psi^{\pm}_{3}}=(\ket{000}\pm\ket{111})/\sqrt{2}$. Here, $\ket{\Psi^{+}_{3}}$ denotes the standard GHZ state, while $\ket{\Psi^{-}_{3}}$ denotes the GHZ state with a relative phase of $\pi$~\cite{NOS21}.

The Lindblad master equation is
\begin{align}
	\label{eq.GHZmaster}
	\dot{\rho}=\mathcal{L}_{3}\rho
	=-i[H_{3},\rho]
	+\sum_{j=1}^{3}\left(P_j\rho P_j^{\dagger}-\frac{1}{2}\{P_j^{\dagger}P_j,\rho\}\right).
\end{align}
Here
$H_{3}
=-J_x(\sigma_{x}\otimes\sigma_{x}\otimes\sigma_{x})
-B(\sigma_1^{z}+\sigma_2^{z}+\sigma_3^{z})$,
where $J_x>0$ is the coupling strength, $B$ is the external field, and
$P_j=\sqrt{\gamma_G}\sigma_j^-$
is the quantum jump operator acting on the $j$th qubit, with $\gamma_G$ the dissipation rate.

Starting from the maximally mixed state $\mathbb{I}/8$, we evolve the system over one cycle along a rectangular encircling loop in the $(\gamma_G,B)$ parameter space with $T=1500$ and $J_x=1$ (see Appendix A). As shown in Fig.~\ref{fig3}(a), the CW encircling drives the maximally mixed state toward $\ket{\Psi^{+}_{3}}$, whereas CCW encircling prepares $\ket{\Psi^{-}_{3}}$ (Fig.~\ref{fig3}(b)).
The measured final-state fidelities reach $0.9191\pm0.0015$ for $\ket{\Psi^{+}_{3}}$ under CW encircling (Fig.~\ref{fig3}(c)) and $0.9130\pm0.0024$ for $\ket{\Psi^{-}_{3}}$ under CCW encircling (Fig.~\ref{fig3}(d)).

Similar to the two-qubit case, the three-qubit entangled-state generation remains robust against dephasing and random perturbations (see Appendix C for details), as respectively illustrated by the dashed and dash-dotted curves in Figs.~\ref{fig3}(a) and \ref{fig3}(b). Hence, our scheme is extendable to the robust generation of multi-qubit  entanglement.

{\it Conclusion---}
We have experimentally demonstrated robust chiral entangled-state generation in the dissipative dynamics of photons.
The observed chiral dynamics is based on a careful design of the Liouvillian, combining steady-state engineering and adiabatic transfer along different sectors of the chiral path.
While the chiral entangled-state generation is apparently similar to the chiral state transfer near EPs, it is based on entirely different mechanism and considerations, and, in our case, only reduces to the non-Hermitian EP encircling in the no-click limit (see  Appendix D for details). Crucially, our scheme does not involve  post selection, making it more adaptable to practical applications in quantum information.
Our work also suggests parametric chiral dynamics as a promising route toward the preparation of many-body states that are useful for quantum simulation,  quantum computation, and quantum metrology.

\begin{acknowledgements}
{\it Acknowledgments---}This work has been supported by the National Key R$\&$D Program of China (Grant No. 2023YFA1406701), the National Natural Science Foundation of China (Grants No. 92265209, No. 12374479, No. 12504413, No. 12474352, No. 92476106,
and No. 12305008). WY acknowledges support
from the Innovation Program for Quantum Science and Technology (Grant No. 2021ZD0301205). HXG. acknowledges support from the China Postdoctoral
Science Foundation (Grants No. 2024M760425 and No. BX20250174). KKW and LX acknowledge support from Beijing National Laboratory for Condensed Matter Physics (No. 2024BNLCMPKF010). Y.H. acknowledges support through the ERC Consolidator project MATHLOCCA (Grant No. 101170485).
\end{acknowledgements}


\appendix
\onecolumngrid
\twocolumngrid

{\it Appendix A: Encircling path in the parameter space---}For the two-qubit case, we consider a rectangular encircling loop in the $(\gamma,\delta)$ plane, parameterized by
\begin{widetext}
	\begin{equation}
		\label{eq.trajectory}
		\gamma(t) =
		\begin{cases}
			\gamma_0,\\
			\gamma_0+\Delta \gamma\dfrac{t-2T_0}{T_0},\\
			\gamma_0+\Delta \gamma,\\
			\gamma_0+\Delta \gamma\dfrac{8T_0-t}{T_0},\\
			\gamma_0,
		\end{cases}
		\qquad
		\delta(t) =
		\begin{cases}
			\pm\dfrac{\Delta \delta }{2T_0}t, & 0<t\le 2T_0,\\
			\pm\Delta \delta, & 2T_0<t\le 3T_0,\\
			\pm\left(\Delta \delta-\dfrac{\Delta \delta(t-3T_0)}{2T_0}\right), & 3T_0<t\le 7T_0,\\
			\pm(-\Delta \delta), & 7T_0<t\le 8T_0,\\
			\pm\left(-\Delta \delta+\dfrac{\Delta \delta(t-8T_0)}{2T_0}\right), & 8T_0<t\le T.
		\end{cases}
	\end{equation}
\end{widetext}
Here $T_0=T/10$, and $+$ ($-$) denotes the CW (CCW) encircling. The parameters $\gamma_0$, $\Delta\gamma$, and $\Delta\delta$ specify the center and boundaries of the loop. In the experiment, we set $\gamma_0=0$, $\Delta\delta=0.04$, $\Delta\gamma=0.18$, and $T=1500$.

For the three-qubit case, we adopt the same rectangular encircling protocol in the corresponding parameter spaces $(\gamma_G,B)$. The loop is parameterized in the generic form
\begin{widetext}
	\begin{equation}
		\label{eq.trajectory_general}
		\gamma_G(t)
		\begin{cases}
			0,\\
			\Delta \gamma_G\dfrac{t-t_1}{t_2-t_1},\\
			\Delta \gamma_G,\\
			\Delta \gamma_G-\Delta \gamma_G\dfrac{t-t_3}{t_4-t_3},\\
			0,
		\end{cases}
		\qquad
		B(t) =
		\begin{cases}
			\pm\Delta B\dfrac{t}{t_1}, & 0<t\le t_1,\\
			\pm\Delta B, & t_1<t\le t_2,\\
			\pm\left(\Delta B-\dfrac{\Delta B(t-t_2)}{t_3-t_2}\right), & t_2<t\le t_3,\\
			\pm(-\Delta B), & t_3<t\le t_4,\\
			\pm\left(-\Delta B+\dfrac{\Delta B(t-t_4)}{T-t_4}\right), & t_4<t\le T.
		\end{cases}
	\end{equation}
\end{widetext}
Here $t_1=T/5$, $t_2=2T/5$, $t_3=3T/5$, and $t_4=4T/5$. The experimental values are $\Delta\gamma_G=1$, and $\Delta B=5$.

{\it Appendix B: Fidelity and concurrence---}We characterize the generated states using the fidelities with respect to the target entangled states. For an $\mathcal{N}$-qubit system, the fidelity is defined as
$\mathcal{F}^{\pm}_{\mathcal{N}}(t)=\mathrm{Tr}\!\left[\ket{\Psi^{\pm}_{\mathcal{N}}}\bra{\Psi^{\pm}_{\mathcal{N}}}\rho(t)\right]$,
where the subscript $\mathcal{N}$ denotes the number of qubits. For the two-qubit case, $\ket{\Psi^{\pm}_{2}}=(\ket{01}\pm\ket{10})/\sqrt{2}$ are the Bell states. For the three-qubit case, $\ket{\Psi^{\pm}_{3}}=(\ket{000}\pm\ket{111})/\sqrt{2}$ correspond to the GHZ state $(+)$ and the GHZ state with a relative phase of $\pi$  $(-)$, respectively. We further quantify the entanglement of the two-qubit states using the concurrence~\cite{W98},
$\mathcal{C}(\rho)=\max\{0,\lambda_1-\lambda_2-\lambda_3-\lambda_4\}$,
where $\lambda_i$ are the eigenvalues, in descending order, of
$\sqrt{\sqrt{\rho}\tilde{\rho}\sqrt{\rho}}$, with
$\tilde{\rho}=(\sigma_y\otimes\sigma_y)\rho^*(\sigma_y\otimes\sigma_y)$. Here, $\rho^*$ denotes the complex conjugate of $\rho$. The concurrence ranges from $0$ for separable states to $1$ for maximally entangled states.

{\it Appendix C: Robustness  analysis for the three-qubit case---}For the three-qubit case, we test the robustness of the protocol against dephasing and random perturbations. In the presence of dephasing, the dynamics are governed by
\begin{align}
	\label{eq.GHZdephasing}
	\dot{\rho}=\mathcal{L}_{3}\rho+\sum_{j=1}^{3}\left(P_j^d\rho P_j^{d\dagger}-\frac{1}{2}\{P_j^{d\dagger}P_j^d,\rho\}\right),
\end{align}
where $P_j^d=0.01\,\sigma_z^j$. The experimental results are shown as the dashed curves in Figs.~\ref{fig3}(a) and \ref{fig3}(b), which remain close to the noiseless results.

We also include random perturbations of the form
$H^{r}_{3}=0.01\,(A_{3}+A_{3}^{\dagger})/2$,
where $A_3$ is an
$8\times8$ random complex matrix whose entries are independently sampled as $a+ib$, with $a$ and $b$ uniformly distributed in $[-1,1]$.  The experimental results are shown as the dash-dotted curves, which remain close to the noiseless case. The shaded regions indicate the range of outcomes from $10$ independent realizations of random perturbations, which exhibit only a small variation.

\begin{figure}
	\includegraphics[width=0.45\textwidth]{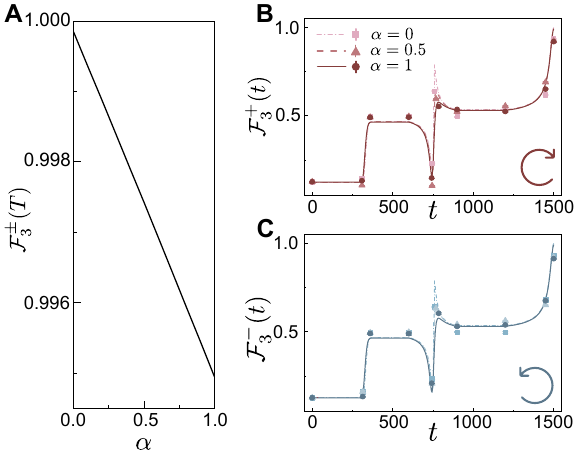}
	\caption{
		Chiral generation of GHZ states under the hybrid Liouvillian formalism.
		(a) Final-state fidelity $F^{+}_{3}(T)$ ($F^{-}_{3}(T)$) with respect to $\ket{\Psi_3^+}$ ($\ket{\Psi_3^-}$) under CW (CCW) encircling as a function of the post-selection strength $\alpha$.
		(b), (c) Time evolution of the fidelities with respect to $\ket{\Psi_3^+}$ and $\ket{\Psi_3^-}$ under CW encircling in (b) and CCW encircling in (c), for three representative post-selection strengths $\alpha=0$, $0.5$, and $1$.
		The other parameters are the same as  those used in Fig.~\ref{fig3}.
	}
	\label{fig4}
\end{figure}
{\it Appendix D: Chiral entanglement generation under the hybrid Liouvillian formalism---}To clarify the connection and distinction of our scheme with the non-Hermitian EP encircling,
we study chiral entanglement generation in the three-qubit system, but with the dynamics governed by a hybrid Liouvillian~\cite{MMC20}
\begin{align}
	\label{eq.GHZhy}
	\dot{\rho}=\mathcal{L}_3(\alpha)\rho=-i(H_{\mathrm{eff}}^{3}\rho-\rho H_{\mathrm{eff}}^{3\dagger})
	+\alpha\sum_{j=1}^{3} P_j\rho P_j^{\dagger}.
\end{align}
Here $H_{\mathrm{eff}}^{3}=H_{3}-\frac{i}{2}\sum_{j=1}^{3}P_j^{\dagger}P_j$, and the other operators are the same as those in Eq.~(\ref{eq.GHZmaster}). The parameter $\alpha\in[0,1]$ characterizes the post-selection strength: $\alpha=0$ corresponds to the no-click limit, where the dynamics is driven by the non-Hermitian Hamiltonian $H_{\mathrm{eff}}^{3}$, while $\alpha=1$ recovers the full Liouvillian dynamics.

Figure~\ref{fig4}(a) shows the numerically simulated final-state fidelities $\mathcal{F}^{+}_{3}(T)$ and $\mathcal{F}^{-}_{3}(T)$ as functions of $\alpha$ under CW and CCW encircling, respectively. As $\alpha$ increases from $0$ to $1$, the fidelity decreases monotonically, mainly because weaker post selection leads to a reduced state purity~\cite{SZ13}. Nevertheless, the fidelity remains high over the entire range of $\alpha$, demonstrating that chiral entanglement generation persists throughout the hybrid-Liouvillian regime.

Experimentally, we choose three representative values, $\alpha=0$, $0.5$, and $1$. Starting from the maximally mixed state $\mathbb{I}/8$, the system evolves along the same three-qubit encircling loop (see Appendix A). The corresponding results for CW and CCW encircling are shown in Figs.~\ref{fig4}(b) and \ref{fig4}(c). For all three values of $\alpha$, CW encircling drives the system toward $\ket{\Psi_3^+}$, while CCW encircling prepares $\ket{\Psi_3^-}$.

These results show that our chiral entanglement-generation mechanism applies not only to the full Liouvillian dynamics, but also to the entire hybrid-Liouvillian family with arbitrary post-selection strength. In the no-click limit $\alpha=0$, the dynamics reduces to the conventional non-Hermitian evolution, and our scheme becomes closely related to chiral state transfer near a non-Hermitian EP~\cite{UMM11,CHY17,QAG26}. In this limit, a non-Hermitian EP appears at $\{B=0,\gamma_G=4/3\}$, and the chiral state transfer is governed by the Riemann-sheet topology of the complex eigenvalue spectrum. However, once the quantum-jump terms are included with finite $\alpha$, the physical mechanism is no longer simply EP encircling. Instead, as discussed above, the chiral behavior results from the designed Liouvillian structure, which combines steady-state selection with adiabatic transport along different sectors of the parameter path. This distinction can be further highlighted by considering a circular parameter trajectory (see more details in Sec. S4 of the Supplemental Material). Such a trajectory still encircles the non-Hermitian EP in the conditional limit, but it does not satisfy the conditions required by our general Liouvillian-based scheme.

\end{document}